\documentclass{mn2e}
\usepackage{graphicx,times}

\def \sanch{S\'anchez-Bl\'azquez}

\def \kms {${\rm{km}\,\rm{s}^{-1}}$}

\def \eza   {{\sc EZ-ages}}

\title[Abundance ratios in red-sequence galaxies]
{\vskip -4mm Abundance ratios in red-sequence galaxies over a wide mass range:
The ``X-planes'' for magnesium, calcium, carbon and nitrogen\vskip -5mm}
\author[Russell J. Smith et al. ]
{Russell J. Smith$^{1}$\thanks{Email: russell.smith@durham.ac.uk}, 
John R. Lucey$^{1}$,
Michael J. Hudson$^{2}$, 
Terry J. Bridges$^{3,4}$ 
\\
$^1$Department of Physics, University of Durham, Durham DH1 3LE\\
$^2$Department of Physics and Astronomy, University of Waterloo, 200 University Avenue West, Waterloo, Ontario N2L 3G1, Canada\\
$^3$Australian Gemini Office, Anglo-Australian Observatory, PO Box 296, Epping NSW 1710, Australia\\
$^4$Department of Physics, Engineering Physics \& Astronomy, Queen's University, Kingston, Ontario K7L 3N6, Canada\\
}

\pagerange{\pageref{firstpage}$-$\pageref{lastpage}} \pubyear{2007}

\begin{document}

\label{firstpage}

\maketitle

\begin{abstract}
We analyse the abundance ratios of the light elements Mg, Ca, C and N, relative to Fe, for 147 red-sequence galaxies in the 
Coma cluster and the Shapley Supercluster. The sample covers a six-magnitude range in luminosity, from 
giant ellipticals to dwarfs at $M^\star+4$.  
We exploit the wide mass range to investigate systematic trends in 
the abundance ratios Mg/Fe, Ca/Fe, C/Fe and N/Fe. 
We find that each of these ratios can be well modelled using two-parameter relations of the 
form ${\rm [X/Fe]} = a_0 + a_1\log\sigma + a_2 {\rm [Fe/H]}$, where $\sigma$ is the velocity dispersion. 
Analysing these ``X-planes'' reveals new structure in the abundance patterns, beyond the traditional 
one-parameter correlations (Mg/Fe--$\sigma$, etc). 
The X-planes for the $\alpha$ elements, Mg and Ca, indicate a positive correlation with velocity dispersion, 
and simultaneously an anti-correlation with Fe/H (i.e. $a_1>0$ and $a_2<0$). 
Taking both effects into account dramatically reduces the scatter, 
compared to the traditional X/Fe$-\sigma$ relations. For C and N, a similar correlation with velocity dispersion 
is recovered, but there is no additional dependence on Fe/H (i.e. $a_1>0$ and $a_2\approx0$).
The explicit dependence of X/Fe on two parameters is evidence that at least two  physical processes are at work in setting 
the abundance patterns. 
The Fe/H dependence of Mg/Fe and Ca/Fe, at fixed $\sigma$, may result from different durations of star formation, from galaxy to galaxy, 
leading to a range in the degree of SNIa contribution to the chemical enrichment. The absence of corresponding Fe/H dependence for 
C and N is consistent with these elements being generated in intermediate and low-mass stars, on time-scales similar to the Fe production.
The $\sigma$ dependence, at fixed Fe/H, is similar for elements Mg, C and N, despite their likely origin in stars of very different 
masses and lifetimes. Ca/Fe is positively correlated with $\sigma$, at fixed Fe/H, but its dependence is significantly less steep than that
of Mg, C and N. We discuss possible origins for this pattern of trends and find no simple prescription that reproduces the observations. 
In particular, if C and N are produced on time-scales comparable to that of Fe, then the X/Fe$-\sigma$ trends cannot be explained solely by a systematic
variation of star-formation time-scale with $\sigma$. 
\end{abstract}

\section{Introduction}

As galaxies evolve, each generation of stars releases newly synthesised elements for incorporation 
into later populations. The final pattern of elemental abundances depends on many factors such as the
star-formation history, the stellar initial mass function (IMF), the yields from different masses of stars in 
supernovae (SNe) and stellar winds, the efficiency of galactic winds in removing enriched material 
from the galaxy, and the inflow of pristine or pre-enriched material from an external reservoir. 
Sophisticated chemical evolution models have been developed using proposals for these ingredients, 
which are all uncertain to some degree (e.g. Matteucci 2001, Gibson et al. 2003 and references therein). 

In our Galaxy and other resolved Local Group systems, the models are constrained by the run of abundance ratios (e.g. X/Fe for element X)
with metallicity Fe/H\footnote{We use the term metallicity interchangeably with iron abundance Fe/H throughout the paper. We will never
refer to ``total'' metallicity, since it is dominated by O, the abundance of which is poorly constrained for external galaxies.}, measured for individual stars from high-resolution
spectra (McWilliam 1997).
Chemical evolution modelling for a wide variety of formation histories is desirable to test the model ingredients
over the full range of conditions relevant to galaxy formation. For example, if the IMF is different in massive starbursts
than in quiescent star-forming environments (e.g. as proposed by Baugh et al. 2005), this should be revealed in different enrichment patterns. 
Since not all classes of galaxy are represented among the systems
amenable to resolved stellar spectroscopy, with early-type galaxies notably absent, it is desirable also to compare models with 
data for more distant galaxies, where only luminosity-weighted average abundances can be measured. 

Age, metallicity and abundance ratio measurements for unresolved galaxies can be obtained from 
intermediate-resolution spectral indices, carefully calibrated via stellar libraries and isochrone synthesis models
(e.g. Worthey et al. 1994; Vazdekis et al. 1999; Thomas, Maraston \& Bender 2003a; Schiavon 2007). 
A key milestone in developing such models was the 
inclusion of non-solar abundances of Mg (or more generally the $\alpha$ elements), first by Tantalo, Chiosi \& Bressan (1998) and
Trager et al. (2000a) and then in the widely-used models of Thomas and collaborators (Thomas et al. 2003a; 
Thomas, Maraston \& Korn 2004). The same basic machinery has since been generalized to allow for more 
flexible elemental mixtures (e.g. Thomas, Maraston \& Bender 2003b; Schiavon 2007). 
Among the other light elements which are readily measurable in unresolved spectra are Ca, C and N (e.g. Graves \& Schiavon 2008). 
Serven, Worthey \& Briley (2005) argue that many more elements could be measured from high 
signal-to-noise spectra, using an enlarged set of index definitions. 

These analysis techniques were initially applied to samples comprised mainly of giant elliptical and S0 galaxies
(e.g. Trager et al. 2000b; Kuntschner et al. 2001; Thomas et al. 2005). Later work extended the coverage to a wider range
of luminosity (e.g. Caldwell et al. 2003; Nelan et al. 2005; Smith, Lucey \& Hudson 2007). The increased baseline helped to reveal
the overall scaling relations of age, metallicity and Mg/Fe, as a function of velocity dispersion, $\sigma$. (Stellar population parameters are generally more
tightly related to $\sigma$ than to other mass proxies such as luminosity.)
On average, less massive galaxies are younger, less metal-rich and have lower  Mg/Fe, than the giant ellipticals. 
The Mg/Fe$-\sigma$ trend has been interpreted as evidence for a shorter time-scale of 
star formation in high-mass galaxies, preventing the incorporation of Fe-rich ejecta from Type Ia supernovae (SNIa), which 
are delayed relative to star formation, into the stars observed today (Thomas et al. 2005). 
Other scenarios have been discussed, e.g. by Trager et al. (2000b), including an IMF which varies as a function of galaxy mass, altering the
relative number of SNIa and SNII, or selective galactic winds to modulate the effective yields.

For other light elements, the picture is less clear. Ca was expected to be enhanced similarly to Mg, since both are produced in the
triple-$\alpha$ chain and released in Type II supernovae (SNII). Instead, Ca appears to have only 
solar abundance with respect to Fe in giant ellipticals (Cenarro et al. 2003; Thomas et al. 2003b). 
For the abundances of C and N, which are probably produced mainly by less massive stars, somewhat 
contradictory results have been obtained (e.g. Clemens et al. 2006; Kelson et al. 2006; Graves et al. 2007).

In this paper, we analyse abundance ratios measured for 81 dwarf galaxies in the Coma cluster (from Smith et al. 2009), and 66 generally 
more luminous galaxies in the Shapley Supercluster (Smith et al. 2007). 
The key observational improvement over previous work is that abundance
ratios are determined for individual galaxies, rather than ensemble averages, and we probe a very wide range in 
luminosity, velocity dispersion and metallicity. A second advance is the introduction of {\it planar} relationships between
the abundance ratios, the metallicity and the velocity dispersion. 
These new scaling relations are potentially powerful constraints on the 
ingredients of chemical evolution models for unresolved galaxies. 
The outline of the paper is as follows. 
Section~\ref{sec:data} briefly reviews the properties of the two galaxy samples and the data employed, and
describes the fits we use to model the abundance ratio correlations. 
In Section~\ref{sec:abund} we investigate empirically the correlations between the abundance ratios and the luminosity, 
velocity dispersion, age and metallicity of the galaxies. We discuss in turn the behaviour of Mg, Ca, C and N, 
in the context of previous observational work. 
Section~\ref{sec:discuss} combines the information obtained for the four elements, and explores how the results relate to models for chemical evolution.
Our principal conclusions are summarized in Section~\ref{sec:concs}. 

\section{Data and methods}\label{sec:data}

\begin{figure*}
\includegraphics[angle=270,width=175mm]{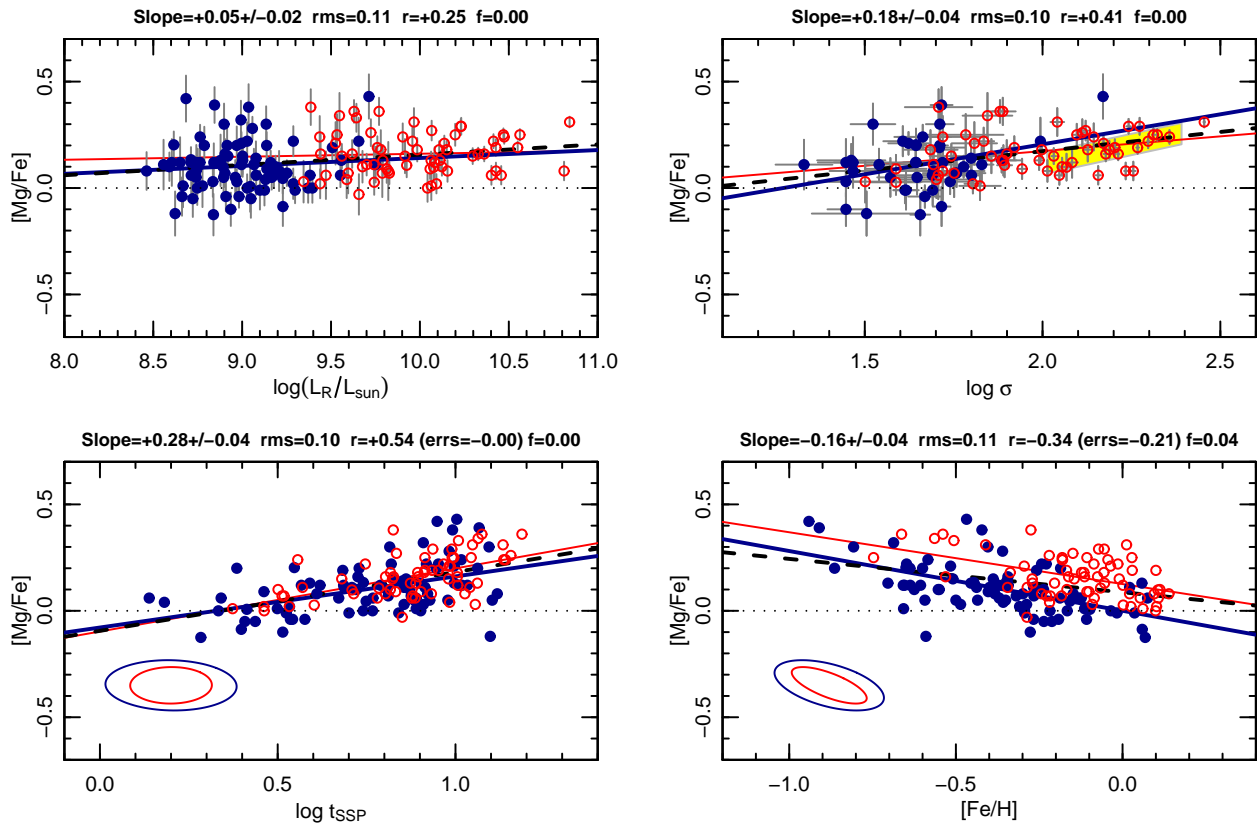}
\caption{Correlations of Mg/Fe with luminosity, velocity dispersion (where available), age and metallicity Fe/H.
The Coma galaxies are shown with blue filled symbols and the Shapley comparison galaxies with red open symbols. 
Indivdual error bars are shown for the luminosity and velocity dispersion panels only; in the age and metallicity panels, we
show a typical error ellipse for each sample in the lower left.
Lines show fits to the combined sample (dashed black), to Coma only (bold blue) and to Shapley only (light red). 
The yellow track in the velocity dispersion panel indicates
the mean trend for giant galaxies, from Graves et al. (2007). The title-bar gives the measured slope, 
and total rms scatter for the fit to the combined sample (see Table~\ref{tab:oneparfits}; for fits to the Coma and Shapley data
separately, see Table~\ref{tab:oneparfitsplit}). 
The product-moment correlation coefficient is $r$. The probability, $f$, of exceeding this $r$-value by chance is computed 
from simulations which include the effect of correlated errors (see Section~\ref{sec:data} for details). 
The $r$ expected from correlated errors alone is given in parentheses.
Age and velocity dispersion are the best single-parameter predictors of the Mg/Fe ratio; a simultaneous fit
prefers a combination of $\sigma$ and Fe/H dependences. Note that the ``offset'' between Shapley and Coma in the Fe/H panel arises because
the Shapley galaxies have much higher $\sigma$ on average.}
\label{fig:mg_fe_corrs}
\end{figure*}

\begin{figure*}
\includegraphics[angle=270,width=175mm]{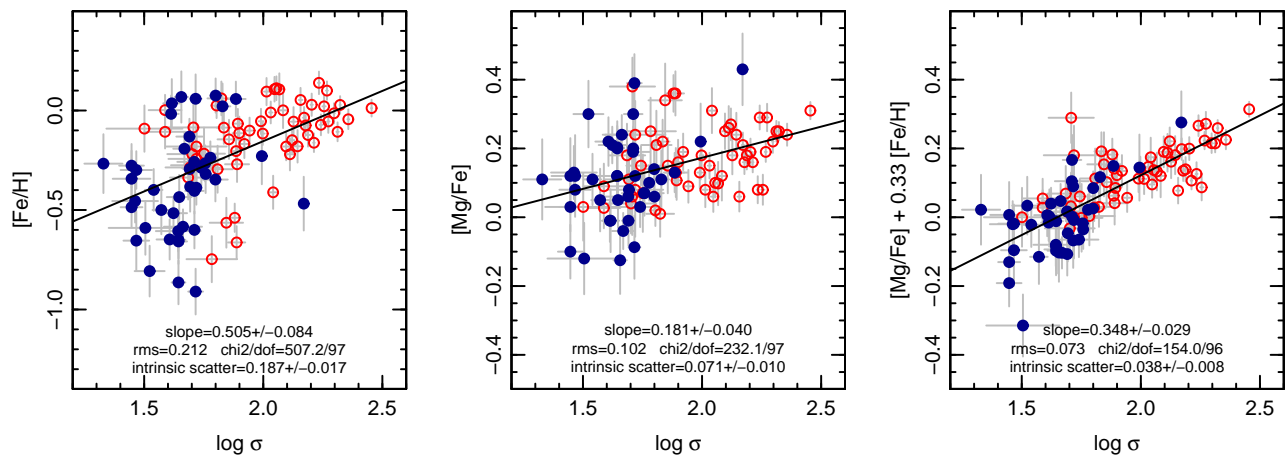}
\caption{
Left: the Fe/H--$\sigma$ correlation for the combined Shapley (red) and Coma (blue) sample. Although these parameters
are correlated, our sample covers sufficient baseline in each parameter that correlations primarily with 
$\sigma$ can be distinguished from correlations primarily with Fe/H. 
Centre: the traditional Mg/Fe--$\sigma$ relation. 
Right: the $\sigma$-projection of the X-plane for Mg. The vertical errorbars account for the anti-correlated errors in Mg/Fe and Fe/H. 
Introducing the additional Fe/H dependence reduces the total rms scatter is from 0.10\,dex to 0.07\,dex (a factor of two in the variance), 
and the intrinsic scatter from 0.07\,dex to 0.04\,dex (a factor of three in variance). 
Note that the reduction of scatter can be appreciated in both the Shapley and the Coma sample 
taken individually. 
} 
\label{fig:mgxplane}
\end{figure*}

\begin{figure*}
\includegraphics[angle=270,width=175mm]{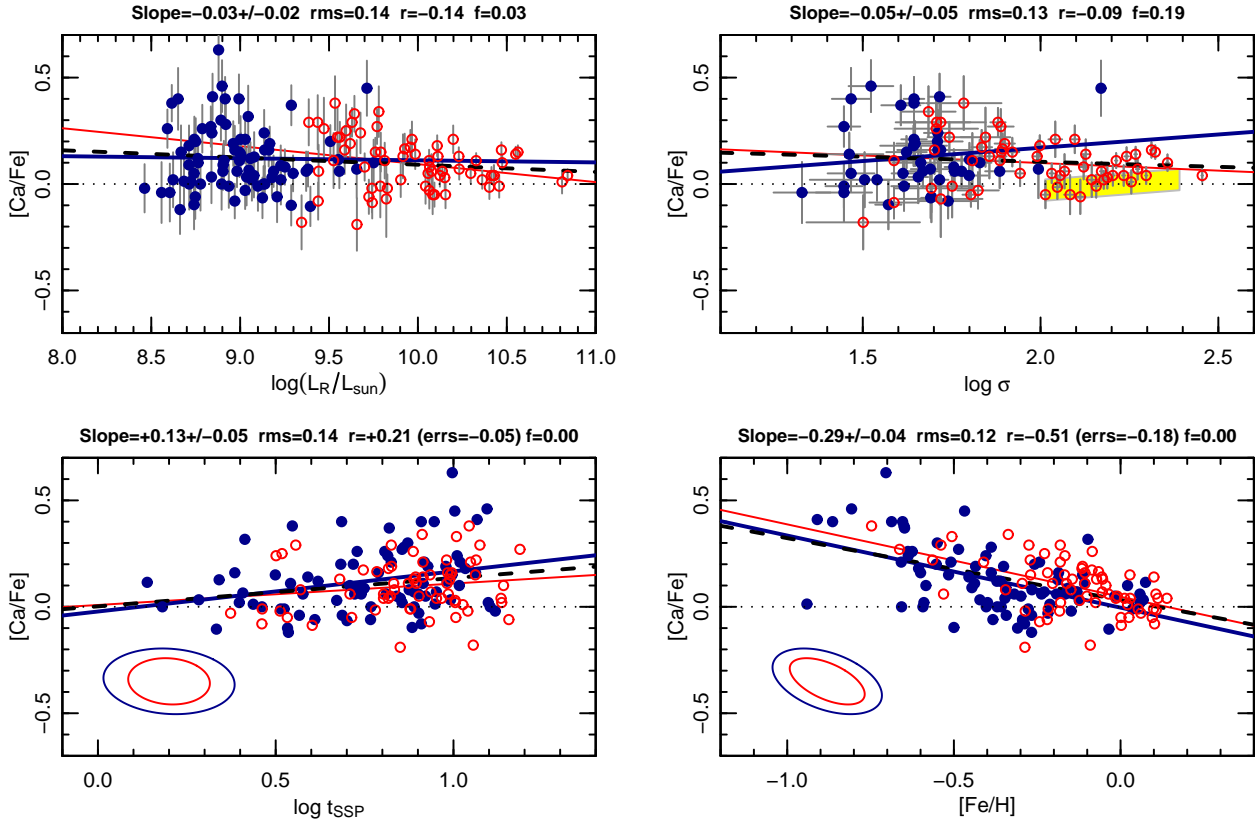}
\caption{Correlations for Ca/Fe. The behaviour is qualitatively 
similar to that of Mg/Fe in Figure~\ref{fig:mg_fe_corrs}, except for the absence of correlation with $\sigma$. 
Symbols and annotations as in Figure~\ref{fig:mg_fe_corrs}.
} 
\label{fig:ca_fe_corrs}
\end{figure*}

\begin{figure*}
\includegraphics[angle=270,width=175mm]{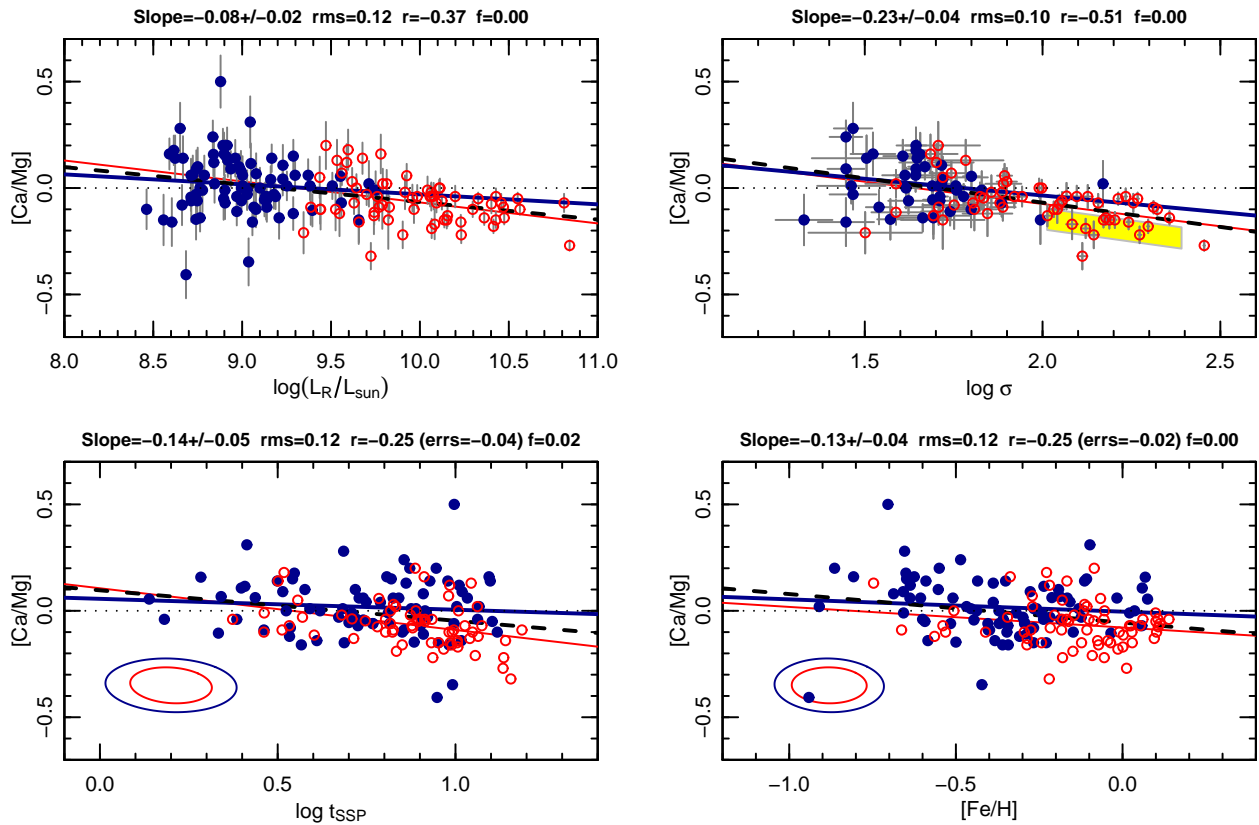}
\caption{Correlations for Ca/Mg. This figure highlights the differences in behaviour for these two $\alpha$ elements, 
in particular the anti-correlation with velocity dispersion. Symbols and annotations as in Figure~\ref{fig:mg_fe_corrs}.} 
\label{fig:ca_mg_corrs}
\end{figure*}

\begin{figure*}
\includegraphics[angle=270,width=175mm]{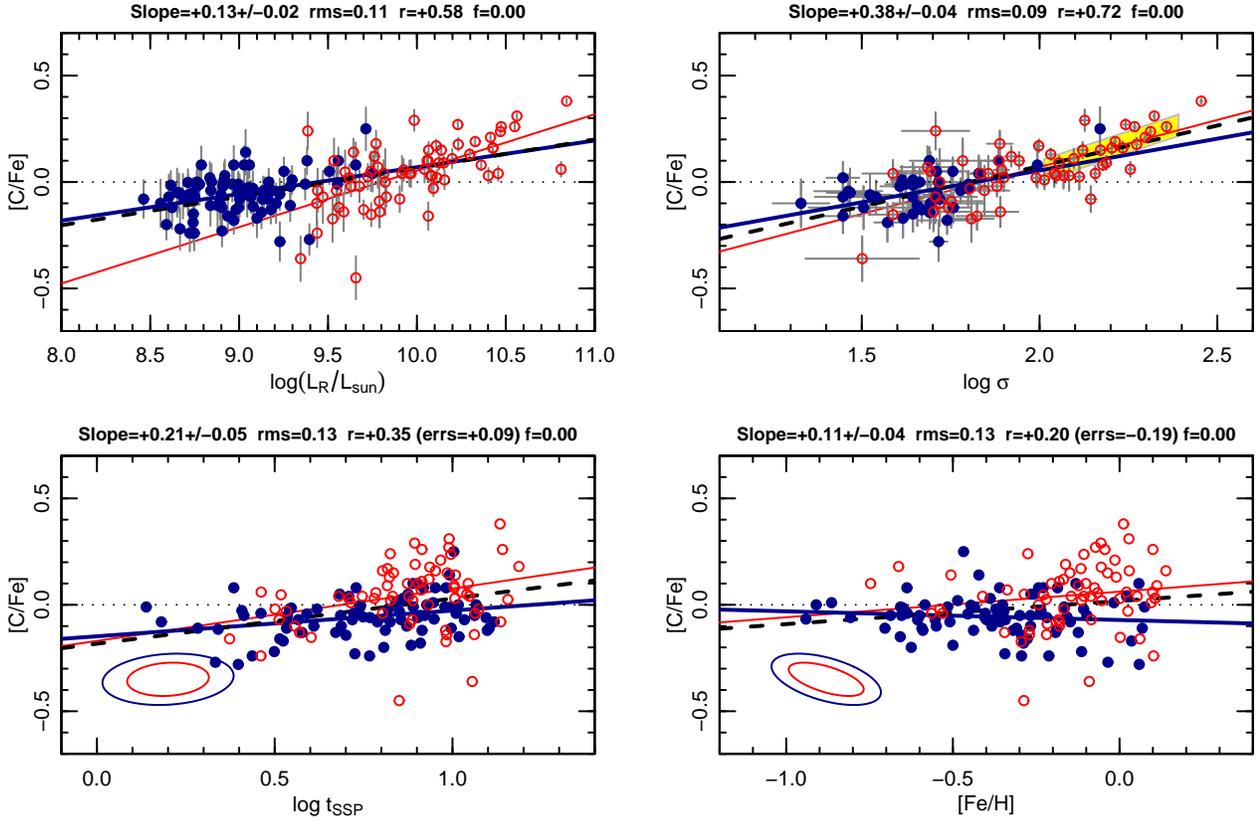}
\caption{Correlations for C/Fe. Note the steep and tight correlation with velocity dispersion, and the absence of
dependence on Fe/H. The latter is quite unlike the case for the $\alpha$ elements 
(Figures~\ref{fig:mg_fe_corrs} and \ref{fig:ca_fe_corrs}). Symbols and annotations as in Figure~\ref{fig:mg_fe_corrs}.}
\label{fig:c_fe_corrs}
\end{figure*}

\begin{figure*}
\includegraphics[angle=270,width=175mm]{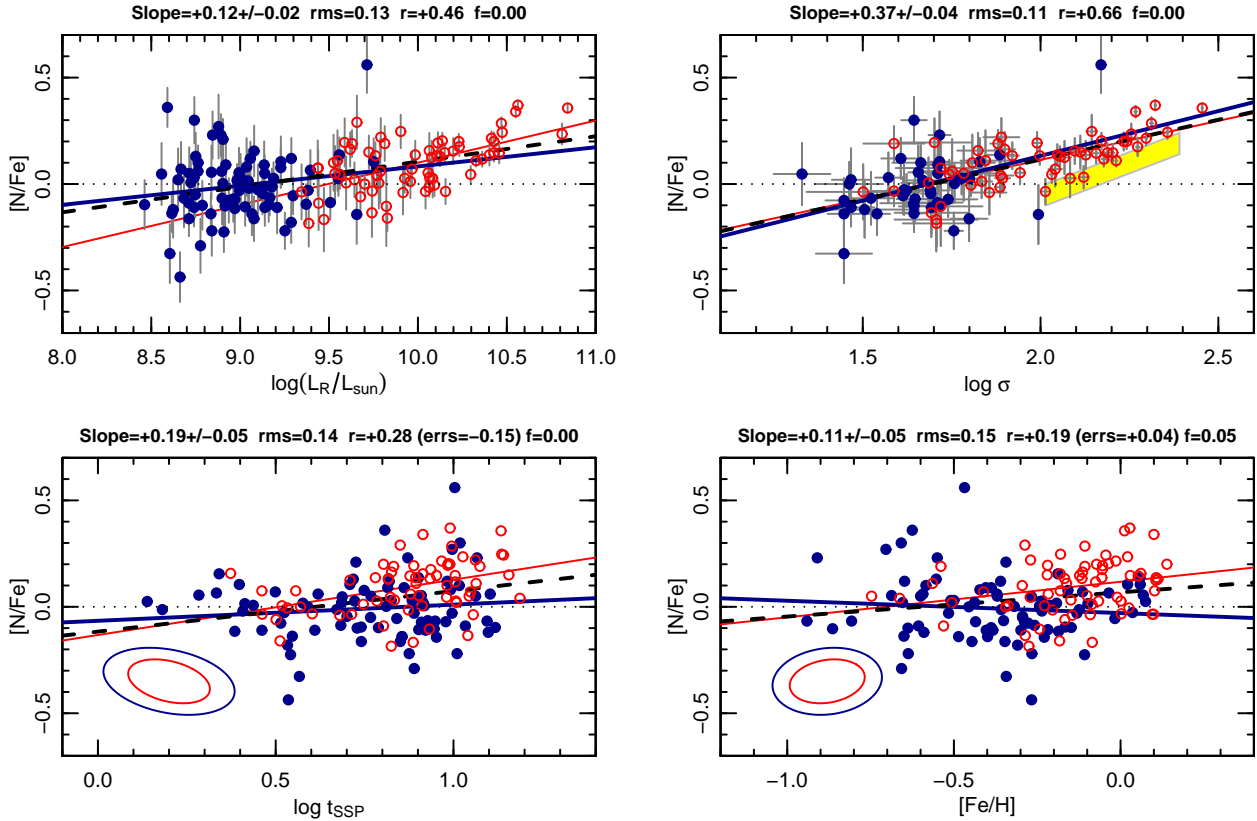}
\caption{Correlations for N/Fe. The trends are strikingly similar to those for C/Fe (Figure~\ref{fig:c_fe_corrs}).
Symbols and annotations as in Figure~\ref{fig:mg_fe_corrs}.
} 
\label{fig:n_fe_corrs}
\end{figure*}

\begin{figure*}
\includegraphics[angle=270,width=175mm]{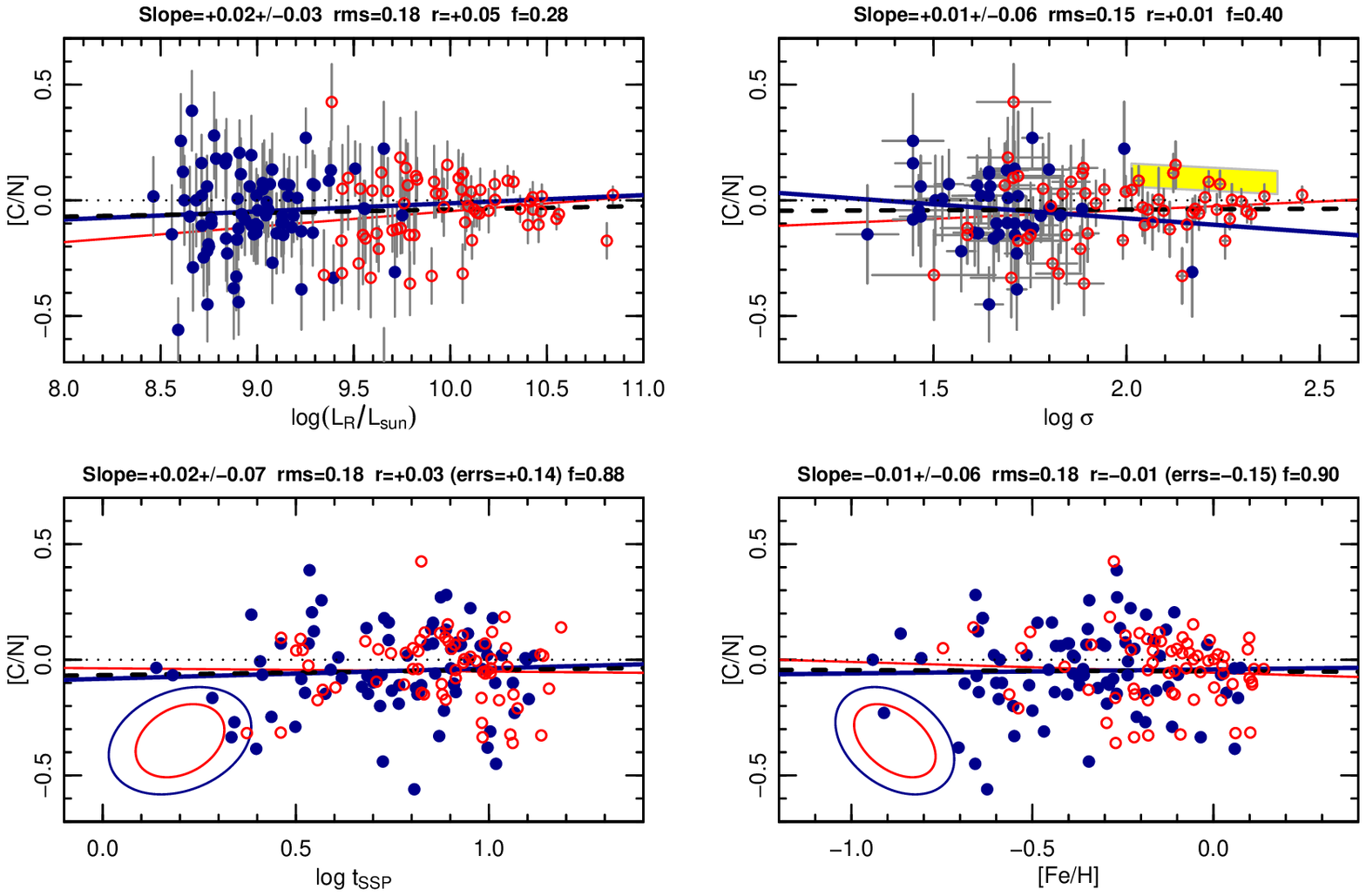}
\caption{Correlations for C/N. The absence of significant trends in this figure emphasizes the systematic similarity in the behaviour of C and N.
Symbols and annotations as in Figure~\ref{fig:mg_fe_corrs}.} 
\label{fig:c_n_corrs}
\end{figure*}

The data used in this paper are based on spectroscopy from Hectospec at the 6.5m MMT for dwarf galaxies in Coma
(Smith et al. 2009), and from AAOmega at the 3.9m Anglo-Australian Telescope for more luminous galaxies in the 
Shapley Supercluster (Smith et al. 2007). The absorption line index data are tabulated in these papers. 
Both studies targeted galaxies in the red sequence, through explicit colour selection (Coma), or post-facto removal of 
galaxies with H\,$\alpha$ emission (Shapley). The treatment of the two datasets is very similar in all respects. 
Velocity dispersion measurements are available for 57 of 66 Shapley galaxies, from the AAOmega observations, 
and for 42 of 81 Coma galaxies, compiled from literature sources. 
The line indices were used to derive stellar population 
parameters by comparison to the models of Schiavon (2007), using the \eza\ code (Graves \& Schiavon 2008). 
The parameters obtained are
SSP-equivalent age, metallicity Fe/H and the light-element abundance ratios Mg/Fe, Ca/Fe, C/Fe and N/Fe. 
The Mg abundance is derived from the Mgb5177 index, Ca from Ca4227, C from C4668 and N from the CN band at 4175\,\AA.
Estimates of the parameter errors and covariances were derived from a Monte-Carlo error model. 
Full details of this process, and tables of parameter estimates, are presented by Smith et al. (2009). 
All quantities used in this paper refer to measurements within an aperture of physical radius 0.95\,kpc 
($\sim$0.5--1.0$\times$ the effective radius, $R_{\rm eff}$) in Shapley, and 0.35\,kpc ($\sim$0.2--0.4$\times R_{\rm eff}$) in Coma. 

Passive galaxies follow many inter-related correlations, such as the so-called ``Z-plane'' (Trager et al. 2000b), which
relates velocity dispersion, SSP-equivalent age and metallicity Fe/H. A fit to the Z-plane for our combined Coma and Shapley sample yields
\[
{\rm [Fe/H]}=(0.69\pm0.07)\log\sigma - (0.70\pm0.09)\log t_{\rm SSP} - 0.90\ , 
\]
where the coefficients are all within $1\sigma$ of the equivalent plane from Trager et al.
Although abundance ratios such as Mg/Fe have traditionally been modelled as functions of $\sigma$, it is not clear a priori whether the 
fundamental dependence is on velocity dispersion, or on age, metallicity or some other related parameter.  
The wide range in galaxy properties spanned by our combined sample will be exploited in this paper to disentangle the 
effects on X/Fe of these mutually-correlated properties. 

We will fit models simultaneously to the two samples, in order to cover the widest baseline in mass and metallicity. The Coma
sample provides excellent coverage of the dwarf galaxy regime, while the Shapley data improve the leverage at high $\sigma$. 
Qualitatively similar results, albeit with larger errors, can be derived from either sample taken individually (see Appendix~\ref{sec:app}).
As a first step, we fit simple correlations of X/Fe with luminosity $L_R$, 
velocity dispersion $\sigma$, SSP-equivalent stellar age $t_{\rm SSP}$ and metallicity Fe/H, considering each of the
latter quantities in turn. These one-parameter tests are summarized in Table~\ref{tab:oneparfits}.

For the correlations with age and metallicity, which are estimated simultaneously with the abundance
ratios in the \eza\ process, it is important to consider the effect of the correlated errors.
To this end we compare the product--moment correlation coefficient $r$ of the observed data against 100 synthetic
datasets for each panel. The simulations assume the ``true'' $X_i$ (e.g. [Fe/H]) are as observed, while the true 
$Y_i$ (e.g. [Mg/Fe]) are zero. The fake ``observed'' data ($x_i$,$y_i$) are drawn from the error ellipse, and the correlation coefficient
recorded. The fraction $f$ of simulations yielding $r$ greater than the observed (one-tailed) is then a measure of the significance which accounts
for the correlated errors. In fact, the errors are sufficiently small that the distribution of galaxies in Fe/H (and age) is well-resolved. For example, 
the observed rms in Fe/H values is 0.24\,dex, while the errors can account for only 0.10\,dex. 
As a result, in all cases where apparently significant trends 
are recovered, the simulation tests confirm that they cannot be generated spuriously by the correlation of errors. 

Considering the one-parameter fits, especially those for Mg (Section~\ref{sec:mgplane}), it became clear
that introducing a second parameter would provide a significantly improved description of the X/Fe variations. 
The two-parameter model with smallest scatter uses the velocity dispersion $\sigma$ and metallicity Fe/H as 
predictors of the X/Fe, in the form: 
\[ 
{\rm [X/Fe]} =  a_0 + a_1 \log\left(\frac{\sigma}{100\,{\rm km\,s}^{-1}}\right) + a_2\, {\rm [Fe/H]} \ . 
\]
These fits are restricted to the 99 galaxies in the combined sample which have velocity dispersion measurements. 
We prefer to employ $\sigma$ rather than luminosity here, because it is not affected by variations in the
stellar mass-to-light ratio, and because in practice tighter correlations are obtained. 
The coefficients and scatters of these relationships, which we call 
the ``X-planes'', are summarized in Table~\ref{tab:xplanefit}. Although we have not explicitly
accounted for the correlated measurement errors in fitting the planes, we do allow for this effect when estimating the 
intrinsic scatter around the X-planes. Moreover, the simulations used to derive the scatter estimates also yield estimates
for the coefficient bias due to the error correlations, which is confirmed to be minimal ($\sim$0.01 for Mg and Ca, 
$\sim$0.04 for C and N).

For completeness, Table~\ref{tab:unifit}) presents the results of more general fits, of the form  
\[
{\rm [X/Fe]} =  b_0 + b_1 \log\left( \frac{\sigma}{100\,{\rm km\,s}^{-1}}\right)  + b_2\, {\rm [Fe/H]} \ + 
\]
\[
~~~~~~~~~~~~~~~~~~~~~~~~~~~~~~~~~~~~~~~~~~~~~~~~~~~~~~~~~~~ b_3 \log\left(\frac{t_{\rm SSP}}{10\,{\rm Gyr}}\right)  + b_4 S\ ,
\]
where the fourth term allows for additional variations with SSP-equivalent age, $t_{\rm SSP}$. 
In the fifth term, $S=1$ for Shapley galaxies and $S=0$ for Coma galaxies. Its coefficient $b_4$ is treated as a nuisance 
parameter to account for unknown offsets between the two data sources, including aperture effects. 
We do not include luminosity as a further predictor here, since it is highly redundant with $\sigma$. (Experiments
with including luminosity do not suggest any significant extra dependences.)
In general, we find that the coefficients $b_3$ and $b_4$ above are consistent with zero, and that using 
the more general fits does not reduce the residual scatter significantly. This justifies, post facto, 
the use of the two-parameter X-planes as a concise description of the abundance correlations. 

Note that the X-plane incorporates velocity dispersion and metallicity on an equal footing
as {\it statistical} predictors of the abundance ratios. However, while $\sigma$ can plausibly be interpreted as 
a {\it causal} predictor of the abundances (i.e. it is an input to chemical evolution models), this is not true for Fe/H, which 
should be an output of any chemical evolution model. Instead, a correlation of X/Fe with Fe/H must indicate a mutual dependence 
of Fe/H and X/Fe on some other aspect of the formation process. We discuss this further in Section~\ref{sec:discuss}.  

\section{Abundance ratio trends}\label{sec:abund}

In this section, we investigate empirically the correlations followed by the abundance ratios Mg/Fe, C/Fe, N/Fe and Ca/Fe. 
We will also make use of some ratios {\it between} the light elements, specifically Ca/Mg and C/N, to isolate similarities
and differences among pairs of elements with similar expected origin.

\subsection{Magnesium}\label{sec:mgplane}

The overabundance of Mg relative to Fe in giant elliptical galaxies has long been recognized 
(e.g. Peterson 1976).
The systematic trends in Mg/Fe (or $\alpha$/Fe, nearly equivalently) 
have been investigated many times, especially since the 
publication of the Thomas et al. (2003a) models
(Trager et al. 2000b; Kuntschner et al. 2001; Thomas et al. 2005; Nelan et al. 2005 and many others). 
A consensus has emerged that Mg/Fe rises 
with velocity dispersion\footnote{A superficially contrary result was obtained by Kelson et al. (2006), who found 
$\alpha$/Fe flat or even decreasing with $\sigma$ in a distant galaxy cluster. 
Note however, that the usual Mgb5177 index was not used in their fits, hence it is not clear which
$\alpha$ elements were being measured, and a direct comparison with Mg/Fe may not be justified.}, 
reaching $\sim$0.3\,dex above solar for giant ellipticals with $\sigma>$200\,\kms.
For dwarf galaxies, the reported Mg/Fe ratios are typically solar (Geha et al. 2003) or modestly super-solar
(e.g. Chilingarian et al. 2008). 
A hint that more complex correlations may be  present was found by \sanch\ et al. (2006), who showed that 
at fixed $\sigma$, older galaxies have higher Mg/Fe on average (their Figure 11).

The single-parameter correlations for Mg/Fe in our sample are presented in Figure~\ref{fig:mg_fe_corrs}. 
We recover the usual increase with velocity dispersion, and a weaker correlation with 
luminosity. The Coma and Shapley samples
seem to follow a common Mg/Fe--$\sigma$ relation, the slope of which is similar to that 
of recent work, including Graves et al. (2007). At the lowest $\sigma$ covered, the average Mg/Fe remains slightly
enhanced over solar (as found by Chilingarian et al. 2008). There is also a correlation with age, $t_{\rm SSP}$, 
which is in fact stronger than the $\sigma$ trend. 
Again, the Coma and Shapley galaxies follow a similar relation. 

A more complex correlation structure is revealed in the panel showing Mg/Fe as a function of Fe/H. Here we might naively 
have expected a positive trend, since high-Fe/H galaxies tend to be more massive, and more massive galaxies
have enhanced Mg/Fe (Figure~\ref{fig:mgxplane}, left and centre). Instead, we recover a significant {\it anti-correlation} of Mg/Fe with Fe/H. (The Monte Carlo simulations
confirm that this result cannot be generated from the correlated errors, at the $>95$\% confidence level.) The Coma and Shapley samples 
follow
parallel but offset relations. Since the Shapley galaxies have higher velocity dispersion on average, this suggests that Mg/Fe 
depends simultaneously on both $\sigma$ {\it and} Fe/H, and that a single-parameter model is insufficient to capture this complex
behaviour. 

The simultaneous fits (Tables~\ref{tab:xplanefit} and \ref{tab:unifit}) confirm that Mg/Fe depends on a linear combination of 
Fe/H and velocity dispersion. On average, a galaxy will have a large over-abundance of Mg if its velocity dispersion
is large, but also if its metallicity is low.  In the X-plane, the slope of the $\sigma$ dependence is much steeper than in 
the single-parameter Mg/Fe$-\sigma$ relation. This is because the high-$\sigma$ galaxies tend to have high Fe/H on average, 
which partially compensates for the higher velocity dispersion. 
The X-plane provides a factor-of-two reduction in variance, compared to 
the  Mg/Fe--$\sigma$ relation (Figure~\ref{fig:mgxplane}, centre and right). 
The observed rms scatter around the X-plane is only $\sim$0.07\,dex, most of which can be accounted for by the estimated errors 
(allowing for correlated errors). The estimated intrinsic component is only $\sim$0.04\,dex. 
The fairly tight (one-parameter) Mg/Fe--$t_{\rm SSP}$ relation seems to be a consequence of the 
X-plane correlation, coupled with the anti-correlation of age and Fe/H at fixed $\sigma$ (the Z-plane of Trager et al. 2000b).
The X-plane is much better predictor of Mg/Fe than the Mg/Fe--$t_{\rm SSP}$ relation, and no
additional age correlation is recovered in the full simultaneous fit. 
The X-plane is also much tighter than a two-parameter model based on $\sigma$ and $t_{\rm SSP}$. 

Note that although our default results are based on the combined Shapley and Coma sample, fitting the two samples separately 
yields consistent coefficients for the X-planes. From each sample we recover significant positive correlation with $\sigma$ and
significant anti-correlation with Fe/H. This confirms that the X-plane is not driven by ``offsets'' between the two samples, 
but is a continuous correlation followed consistently by galaxies in both clusters. The results of these separate-sample fits
are provided in Appendix~\ref{sec:app}.

We conclude that the X-plane provides the most accurate two-parameter description of 
the Mg/Fe data, is much preferred over one-parameter models, and is not significantly improved upon by fits with more
parameters. Motivated by this finding, we continue by applying the same form of model to the other measured abundance ratios. 

\subsection{Calcium}

It has long been suggested (e.g. Vazdekis et al. 1997; Worthey 1998) that the abundance of Ca does not behave identically to that 
of Mg, despite the expectation that the production of both of these $\alpha$-elements should be dominated by SN II. 
Thomas, Maraston \& Bender (2003b) showed that Ca/Mg is subsolar by $\sim$0.2\,dex for
giant ellipticals with $\sigma\ga250$\,\kms, but that Ca/Mg is approximately solar for low-mass galaxies $\sigma\la100$\,\kms, 
including Local Group dwarf spheroidals. 
In the Lick system, Ca is traced by the Ca4227 index\footnote{An alternative tracer is provided by 
the Ca triplet at 8500--8700\,\AA, which has also been used to infer under-abundant Ca, and/or variations in the stellar 
initial mass function (Saglia et al. 2002; Cenarro et al. 2003).}, which is sensitive to the neighbouring CN
band-head, as discussed by Prochaska, Rose \& Schiavon (2005).
The \eza\ approach in principle allows for the effect of varying C/Fe and N/Fe, and should help to decouple 
the Ca abundances. Even with this method, however, Graves et al. (2007) report only a weak increase of 
Ca/Fe with $\sigma$ for giant red-sequence galaxies, 
with approximately solar abundance on average, suggesting the anomalous results do not arise from 
CN contamination. 

Our correlations for Ca/Fe are shown in Figure~\ref{fig:ca_fe_corrs}. We confirm the absence of a
significant trend with velocity dispersion, or with luminosity, although our average Ca/Fe ratios are slightly super-solar throughout. 
There is however a strong anti-correlation with Fe/H, which is very similar to the case for Mg/Fe,
and a weak correlation with age is also recovered. 

The X-plane and multi-parameter fits confirm that the dominant dependence for Ca/Fe is an anti-correlation with Fe/H. 
With the simultaneous fits, we do recover an increase of Ca/Fe with $\sigma$, at fixed Fe/H. This was ``hidden'' in the single-parameter fits
by the metallicity trend: high-$\sigma$ galaxies tend also to have high Fe/H, causing compensating effects on Ca/Fe.  
In the multi-parameter fit, there is marginal evidence for additional correlation of Ca/Fe with age, 
although this makes insignificant contribution to the total variance.  

To highlight the similarities and differences between Mg and Ca, we show in Figure~\ref{fig:ca_mg_corrs}
the equivalent correlations for Ca/Mg. The strongest trend is an anti-correlation with velocity dispersion. 
The Shapley and Coma samples seem to follow a common relation,
with Ca under-abundant by $\sim$0.1\,dex relative to Mg in galaxies with 
$\sigma\ga50$\,\kms\ or $L_r \ga 10^9 L_\odot$, but near solar in dwarf galaxies. We conclude that Ca indeed behaves somewhat
differently from Mg in the X-plane formulation. However, the relations for Ca and Mg share  similar features when viewed in contrast to 
the very different correlations found for elements C and N which we explore next. 
 
\subsection{Carbon}

There is a long-standing debate as to whether short-lived high-mass stars (via Wolf--Rayet winds) or long-lifetime 
intermediate/low-mass stars (through the planetary nebula phase) are primarily responsible for enrichment of C in the solar 
neighbourhood (e.g. Gustafsson et al. 1999; Carigi et al. 2005). 

Determinations of C/Fe, and its trend with mass, in unresolved galaxies have been somewhat contradictory. 
\sanch\ et al. (2006) report C/Fe rising with $\sigma$, but less steeply than Mg, while
Clemens et al. (2006) find C/H (and therefore also C/Fe) declining with $\sigma$, and Kelson et al. (2006) argue for no strong
$\sigma$ dependence in C/Fe. More recently, 
Graves et al. (2007), using \eza, recover a C/Fe increase with $\sigma$ that is even steeper 
than that of Mg/Fe. 

Figure~\ref{fig:c_fe_corrs} shows the correlations followed by C/Fe in our data. We find that C/Fe increases
with luminosity and with velocity dispersion. The Shapley and Coma galaxies follow a similar correlation
with $\sigma$, where the slope is also consistent that reported by Graves et al. for giant galaxies. 
In the luminosity correlation, by contrast, there is some evidence for a change in slope between giants and dwarfs.
There is a weak increase of C/Fe with age 
(as seen for both Mg/Fe and Ca/Fe). However, the behaviour of C differs strikingly from that of Mg and Ca in the 
metallicity panel: we do not recover a strong anti-correlation of C/Fe with Fe/H, and indeed the correlation
is marginally positive.

The simultaneous fits again confirm the above discussion. The behaviour of C/Fe can be described by a strong underlying
correlation with $\sigma$. After accounting for this, there is a marginally significant anti-correlation with Fe/H, which
was hidden by the $\sigma$ dependence. However, the metallicity trend of C is very 
much shallower than the equivalent trends for Mg and Ca.

\subsection{Nitrogen}

The production of N is thought to be dominated by intermediate-mass stars, through 
mass loss in the asymptotic giant branch (AGB) and planetary nebula phases (van den Hoek \& Groenewegen 1997), 
although there could also be contributions from rapidly-rotating massive stars (Meynet \& Maeder 2002). 
Previous work on early-type galaxies has suggested that N/Fe is enhanced over solar, and that the enhancement increases with $\sigma$. 
\sanch\ et al. (2006) found that to reproduce their observed index--$\sigma$ relations, N/Fe should increase with 
$\sigma$ but less steeply than Mg/Fe.
Kelson et al. (2006) found N/Fe rising steeply with velocity dispersion (converted from their quoted $\alpha$/N and 
$\alpha$/Fe trends). 
Using \eza, Graves et al. (2007) report an increase in N that is much stronger than the equivalent for Mg\footnote{Very recently, 
Toloba et al. 2008 report only a very weak dependence of the NH 3360\,\AA\ band on velocity dispersion, which appears contrary to the
results from the CN band used here and in most other work.}.

Our results for N are presented in Figure~\ref{fig:n_fe_corrs}, and are strikingly similar to the equivalent relations for C: 
there are strong correlations of N/Fe
with luminosity and velocity dispersion, and only weak dependence on age and metallicity. 
Our N/Fe$-\sigma$ slope is similar to that reported by Graves et al. for giant galaxies, but our relation has a 
higher zero-point, by $\sim$0.1\,dex. 

As expected from Figure~\ref{fig:n_fe_corrs}, the coefficients of the X-plane for N are indistinguishable from those
of C. To emphasise the absence of systematic variation between these elements, Figure~\ref{fig:c_n_corrs} shows single-parameter
correlations for the C/N ratio, where no significant slopes are obtained.

\begin{figure}
\includegraphics[angle=270,width=80mm]{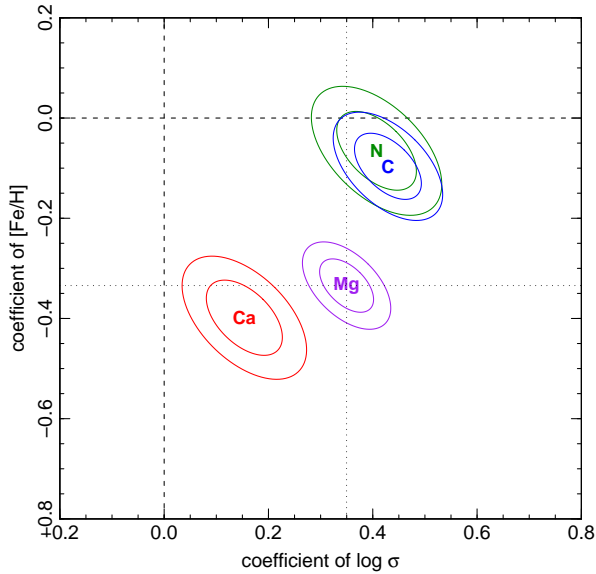}
\caption{Confidence intervals on the parameters of the X/Fe planes. For each element, the ellipses show the 
68\% and 95\% confidence intervals on the two slope coefficients. The dashed lines indicate zero variation, while
the dotted lines coincide with the parameters for the Mg plane.}
\label{fig:modellip}
\end{figure}

\section{Discussion}\label{sec:discuss}

As noted in the previous section, our results for the average trends of X/Fe as a function of 
velocity dispersion largely agree with previous work in the regime of giant galaxies. In particular, 
our (single-parameter) X/Fe$-\sigma$ slopes are consistent with those of Graves et al. (2007), who used
the the same stellar population models (Schiavon 2007) and inversion method (Graves \& Schiavon 2008). 

\begin{figure*}
\includegraphics[angle=0,width=175mm]{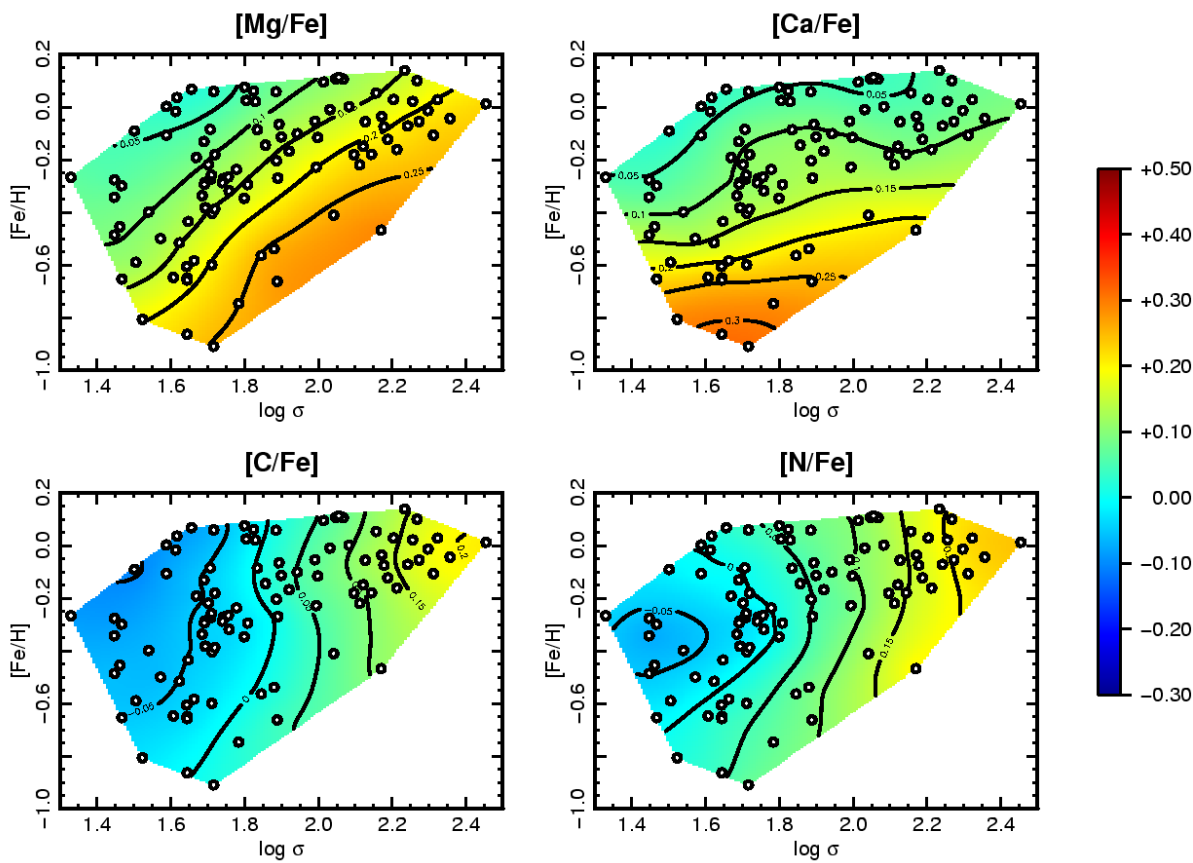}
\caption{The abundance ratios on the Fe--$\sigma$ plane. The points show the Fe/H--$\sigma$ relation for the 
combined Coma and Shapley galaxy sample, as in the left panel of Figure~\ref{fig:mgxplane}. The colour shading and contours
were derived by heavily smoothing the X/Fe ratios onto this plane using a Kriging model (see e.g. Cressie 1993).
Contours are spaced at 0.05\,dex intervals. Note the distinction between C and N, where the near-vertical contour
indicate dependence only on $\sigma$, and Mg and Ca, where the contours are inclined and show a strong dependence 
on Fe/H.}\label{fig:smoox}
\end{figure*}

Where our study breaks new ground is in probing a much larger range in galaxy mass and metallicity, 
and in analysing the abundance patterns using multi-parameter fits. 
The results of our X-plane fits for all four elements are summarized in Figure~\ref{fig:modellip}, in 
the form of confidence intervals on the fit coefficients. 
(An equivalent figure for the separate X-plane fits to Coma and Shapley is shown in Appendix~\ref{sec:app}, and confirms
that the main results can be obtained from either sample individually.)
An alternative way to visualise the bivariate
dependence of X/Fe is by means of contours in the Fe/H--$\sigma$ plane, as shown in Figure~\ref{fig:smoox}. 
The X-planes are readily seen in our data owing to the wide baseline covered in both $\sigma$ and Fe/H. 
These results could not easily have been recovered from earlier samples which were dominated by 
metal-rich high-$\sigma$ giant ellipticals. 

A detailed comparison against realistic chemical evolution models is clearly required to interpret 
the results we have presented. 
In the following sections, we limit the discussion to some qualitative arguments.
First, we argue that variation in star-formation time-scale, at fixed $\sigma$, is responsible for imprinting 
the $\alpha$/Fe-vs-Fe/H trends, while leaving the C/Fe and N/Fe ratios independent of Fe/H (Section~\ref{sec:alpha}).
Second the comparable increase of X/Fe with $\sigma$ for all four light elements requires a separate explanation, since time-scale
and other progenitor-mass effects should not distinguish between SNIa products (Fe) versus AGB products (C and N)
(Section~\ref{sec:sniasupp}). Third, we comment on the apparent discrepancy in the behaviour of Ca compared to Mg 
(Section~\ref{sec:calciumpuzzle}).

\subsection{Fe/H trends: a range in star-formation time-scale at fixed velocity dispersion}\label{sec:alpha}

For individual stars in the solar neighbourhood, the Mg/Fe ratio is enhanced at low Fe/H, where SNII enrichment is important, 
and drifts towards solar values at high Fe/H, as SNIa become dominant (McWilliam 1997 and references therein).
This pattern results from SNIa being on average significantly delayed relative to star formation (and SNII), as expected if their
progenitors are low/intermediate mass stars\footnote{Recent evidence that a fraction of SNIa occur promptly after star formation
(e.g. Sullivan et al. 2006) does not affect this general scenario.}

Of course, in galaxies we observe only an average over the enrichment history, rather than the ``snapshots'' of chemical evolution 
revealed by individual stars. None the less, if other galaxies experienced similar enrichment processes, operating over
a variety of time-scales, we might expect the integrated abundances to follow a qualitatively similar trend. Galaxies
where star formation persisted for longer would have more time to incorporate SNIa ejecta, raising the average Fe/H and lowering
the average Mg/Fe ratio. 

\begin{figure*}
\includegraphics[angle=0,width=175mm]{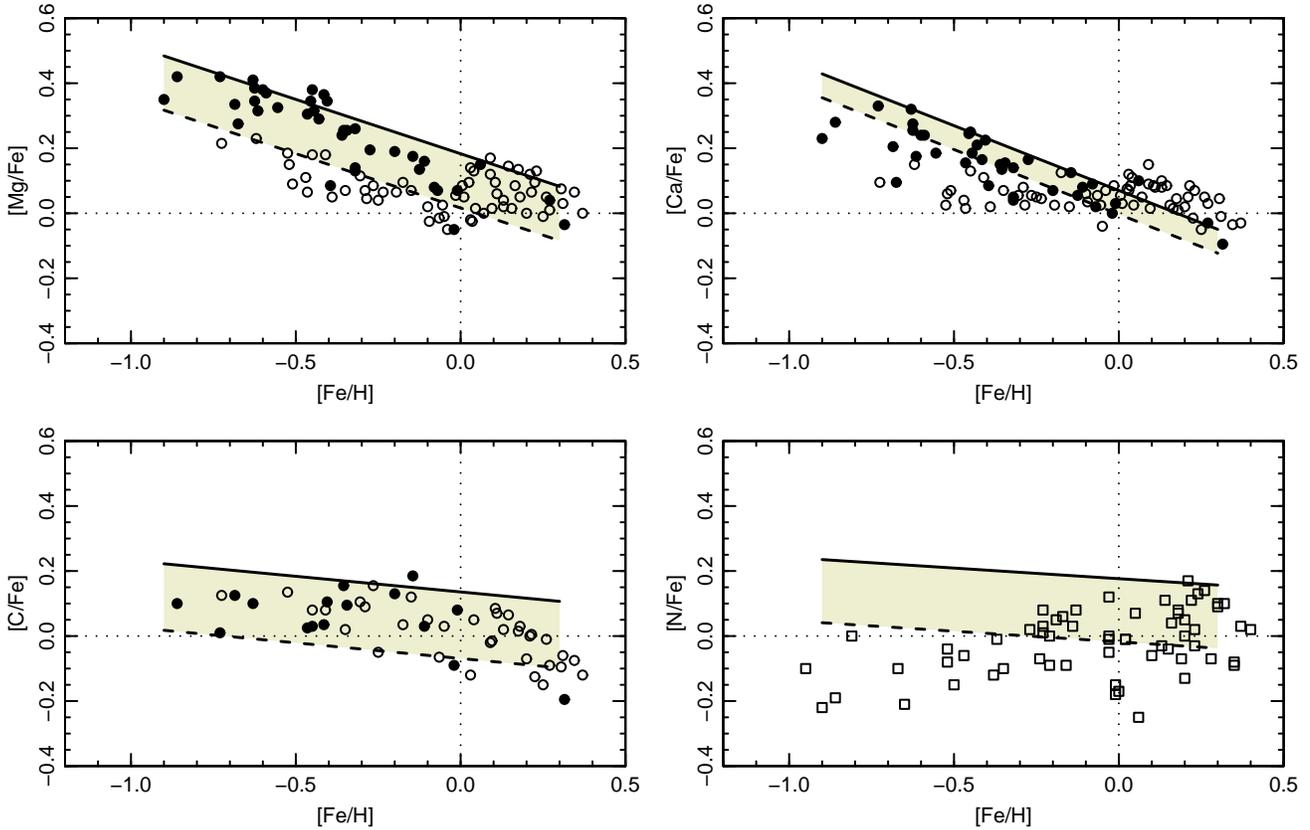}
\caption{Comparing abundance ratio data for milky way stars (points) and passive galaxies (lines and shading). 
The filled and open circles indicate thick and thin disk stars, respectively, from Bensby et al. (2005). In the N/Fe panel, squares show stars from
Ecuvillon et al. (2004) and Israelian et al. (2004). 
In each panel, the X-plane fits (Table~\ref{tab:xplanefit}) have been used to derive the mean X/Fe--Fe/H relation for galaxies of 
velocity dispersion $\sigma=50$\,\kms\ and $\sigma=150$\,\kms\ (dashed and solid black lines, respectively). 
The agreement between the stellar data and the external galaxies suggests a degree of similarity in their enrichment processes.}
\label{fig:bensbycomp}
\end{figure*}

A key puzzle in earlier work on elliptical galaxies was that higher-metallicity galaxies
have higher Mg/Fe ratios, i.e. opposite to the pattern in galactic stars. By analysing the X-plane for Mg, we have shown that there 
{\it is} an anti-correlation like that in the stars, {\it after referencing to a common velocity dispersion}. 
Intriguingly, we find that the slope of the Mg/Fe-vs-Fe/H dependence at fixed $\sigma$ ($-0.33$) agrees
quantitatively with that recovered for galactic disk stars, over a similar metallicity range 
(Figure~\ref{fig:bensbycomp}, top panel). The resemblance is particularly striking for the thick disc stars taken alone. 
We may interpret this slope as the characteristic signature of star-formation being quenched during the transition
from SNII-dominated to SNIa-dominated abundance patterns. (The thin disk under went the same transition at lower metallicity).

For Ca, we found an X-plane which to first order resembles that of Mg, as expected if they are both produced mainly in SNII, and 
in particular, the anti-correlation with Fe/H is similar to that for Mg.
Thus both the $\alpha$ elements qualitatively support a picture where, at a given velocity dispersion, 
different galaxies experienced different time-scales of star formation, leading to an anti-correlation of $\alpha$/Fe with Fe/H. 
Again, the Fe/H coefficient of the X-plane is well matched to the Ca/Fe--Fe/H relation for thick disk stars (Figure~\ref{fig:bensbycomp}, 
second panel).

While it is generally agreed that the $\alpha$ elements Mg and Ca are released in SNII from high-mass stars ($>8$\,$M_{\odot}$), 
there is less consensus on the stars which dominate production of C and N. In the case of C, there is continuing debate as to 
the relative contributions from high-mass ($>$8\,$M_{\odot}$) vs low/intermediate mass stars (1--8\,$M_{\odot}$) (Gustafsson et al. 1999; 
Carigi et al. 2005). For N, there is an additional dimension to the debate, on the relative importance of primary and secondary 
synthesis, i.e. whether yields depend on the metallicity of the gas from which the star was born (e.g. Gavil\'an, Moll\'a \& Buell 2006). 

Recent work on galactic disk stars has shown little if any increase in C/Fe  
and N/Fe with declining Fe/H (Israelian et al. 2004; Bensby \& Feltzing 2006). The simplest interpretation is that 
the production of C and N keeps pace with that of Fe, i.e. these elements are produced in lower-mass stars with similar characteristic
lifetimes as the SNIa progenitors.
Our X-plane results suggest that the production sites of C and N in other galaxies are likewise dominated by low/intermediate-mass 
stars, which would lead naturally to a corresponding independence of C/Fe and N/Fe with respect to Fe/H. 
Figure~\ref{fig:bensbycomp} confirms the similarity of the disk abundance pattern to the X-plane slopes for C. For N, the stellar
data (from a different literature source than the other elements) suggest a mildly positive correlation which is not seen in the
X-plane slope.

\subsection{Trends of X/Fe with velocity dispersion}\label{sec:sniasupp}

Having discussed the variation in X/Fe at fixed $\sigma$, we now consider the systematic trends as a function of $\sigma$.
The key constraint is that the $\sigma$ dependences of Mg, C and N are consistent with being identical, 
with X/Fe\,$\propto${}$\sigma^{0.4}$ at constant Fe/H. (Note that the similarity in the $\sigma$ trends is apparent in the 
X-planes, where the different dependences on Fe/H are accounted for, but not in the traditional single-parameter scaling relations.)

Several different mechanisms have been proposed to explain the traditional Mg/Fe$-\sigma$ correlation, by
altering the relative fractions of SNIa vs SNII enrichment (see discussion by Trager et al. 2000b). 
The most widespread interpretation is that 
galaxies of higher velocity dispersion formed their stars on more rapid time-scales (Thomas et al. 2005), so that SNIa make less
contribution to enriching the stellar population. However, if C and N are produced on nearly the same time-scale as Fe, 
then this scenario predicts that {\it only} the $\alpha$ elements Mg and Ca should be strongly 
correlated with $\sigma$. Thus the steep C/Fe$-\sigma$ and N/Fe$-\sigma$ relations (in combination with the weak metallicity 
dependence of these elements) argues against a systematic variation in star-formation time-scale with velocity dispersion. 

An alternative explanation for the Mg/Fe$-\sigma$ relation invokes variation of the IMF slope in favour of higher-mass stars at high
velocity dispersion. This would produce a higher ratio of SNII enrichment (Mg, Ca) compared to SNIa (Fe), as required, but would not 
increase the C and N abundances, if these element are indeed generated in stars of mass comparable to that of SNIa progenitors. 
Thus the steep  C/Fe$-\sigma$ and N/Fe$-\sigma$ relations also argue against an explanation in terms of IMF variation. 
Selective outflows of gas provide another way to vary the effective yields of SNII and SNIa. However, any mechanism that
operates only in the early SNII-dominated phase will fail to match the observed trends because it will not distinguish between 
Fe production in SNIa and C and N production in AGB stars. For later galactic winds, it is plausible that SNIa products are 
ejected by the explosion energy, while AGB products are retained. 
However, this would lead to enhanced C/Fe and N/Fe in low-$\sigma$ galaxies, opposite to the sense of the observed trend. 
A final possibility is that the number of SNIa explosions (per unit stellar mass) may be smaller in high-$\sigma$ galaxies than in 
low-$\sigma$ objects. This could occur if larger star--star or cloud--cloud relative velocities in a star-forming region somehow 
suppress the formation of binary SNIa progenitors. However, this model would also predict low Fe abundances in the hot interstellar 
gas, which could be in conflict with recent X-ray analyses (Humphrey \& Buote 2006; Tawara et al. 2008). 

In this discussion we have focussed on the X/Fe--$\sigma$ relations, but the Fe/H--$\sigma$ relation itself is also relevant
here. In particular, models in which the Mg/Fe$-\sigma$ relation arises from reduced incorporation of Fe-rich 
SNIa ejecta at high $\sigma$ (e.g. varying time-scales, SNIa rate modulation) would at face value predict lower Fe/H in 
higher-$\sigma$ galaxies, in contradition to the observed positive correlation (Figure~\ref{fig:mgxplane}, left). 
Taken together, the variation of X/Fe and Fe/H with $\sigma$ cannot be reproduced with any of the traditional models
for explaining abundance ratio trends, at least in their simplest form. It is conceivable that combinations of these scenarios, 
e.g. SNIa suppression with an additional non-selective galactic wind, reducing all metals equally in low-$\sigma$ galaxies, might 
be compatible with the observed correlations. Quantitative chemical evolution modelling is clearly required to test more complex 
models of this kind.

\subsection{The calcium puzzle}\label{sec:calciumpuzzle}

Finally, we address the apparently anomalous behaviour of Ca/Fe with respect to velocity dispersion. The coefficient of $\sigma$ in the X-plane for Ca 
is much smaller than the $\sim$0.4 shared by the other three elements. While Ca/Mg is approximately solar at low $\sigma$, the shallower
trend yields surprisingly low Ca/Mg for massive galaxies, as has been noted before (e.g. Thomas et al. 2003b). 

Pipino \& Matteucci (2004) point out that SNIa make a non-negligible contribution to Ca enrichment, according to the yields
they adopt (from Nomoto et al. 1997). This results in reduced separation between the Ca/Fe ratios between SNII-dominated and SNIa-dominated
enrichment. Since the ratios are quoted relative to the solar value, this leads to lower than expected Ca/Fe (i.e. subsolar Ca/Mg),
in the SNII-enriched populations which predominate at high $\sigma$. 
The chemical evolution models of Pipino \& Matteucci (2004) predict [Ca/Mg]\,$\approx$\,--0.15 for massive galaxies, 
in good agreement with the high-$\sigma$ offset in our Figure~\ref{fig:ca_mg_corrs}. Thus scenarios which modulate the 
relative importance of SNIa and SNII as a function of $\sigma$ should, qualitatively at least, produce a shallower 
dependence for Ca/Fe than for Mg/Fe, as observed. However, the same argument would also predict a shallower dependence of 
Ca/Fe on Fe/H, which is not seen in our data. 

An additional anomaly is that alone among the abundance ratios considered here, Ca/Fe exhibits a 
significant dependence on SSP-equivalent age in the extended multiparametric fit (Table~\ref{tab:unifit}). 
Specifically, older galaxies have lower Ca/Fe than younger galaxies with the same $\sigma$ and Fe/H. Table~\ref{tab:unifit} shows that 
correcting to common age yields a $\sigma$ dependence somewhat closer to the value for Mg (0.22$\pm$0.06 versus 0.31$\pm$0.04). 
We do not speculate on the cause of the age correlation here. More detailed chemical evolution modelling is clearly required to resolve the
enduring puzzle of Ca abundances in passive galaxies.

\begin{table}
\caption{The single-parameter fits for X/Fe. The quoted rms is the total observed scatter, i.e. uncorrected for measurement error, in dex.
The table is organised for ready comparison between the four elements.}\label{tab:oneparfits}
\begin{tabular}{lcccccc}
\hline
Ordinate & X & zero-point & slope & rms\\
\hline
$\log\left(\frac{\sigma}{100\,{\rm kms}^{-1}}\right)$ & Mg      & $+0.17\pm0.01$ & $+0.18\pm0.04$ & 0.10 \\
                                                      & Ca      & $+0.10\pm0.02$ & $-0.05\pm0.05$ & 0.13 \\
                                                      & C       & $+0.07\pm0.01$ & $+0.38\pm0.04$ & 0.09 \\
                                                      & N       & $+0.11\pm0.01$ & $+0.37\pm0.04$ & 0.11 \\
\hline
$\log\left(\frac{L_r}{10^9 L_\odot}\right)$  & Mg      & $+0.11\pm0.01$ & $+0.05\pm0.02$ & 0.11 \\
                                             & Ca      & $+0.12\pm0.01$ & $-0.03\pm0.02$ & 0.14 \\
                                             & C       & $-0.07\pm0.01$ & $+0.13\pm0.02$ & 0.11 \\
                                             & N       & $-0.01\pm0.01$ & $+0.12\pm0.02$ & 0.13 \\
\hline
$\log\left(\frac{t_{\rm SSP}}{10\,{\rm Gyr}}\right)$ & Mg      & $+0.18\pm0.01$ & $+0.28\pm0.04$ & 0.10 \\
                                                & Ca      & $+0.14\pm0.02$ & $+0.13\pm0.05$ & 0.14 \\
                                                & C       & $+0.03\pm0.01$ & $+0.21\pm0.05$ & 0.13 \\
                                                & N       & $+0.07\pm0.02$ & $+0.19\pm0.05$ & 0.14 \\
\hline
$\left[{\rm Fe/H}\right]$ & Mg      & $+0.09\pm0.01$ & $-0.16\pm0.04$ & 0.11 \\
                          & Ca      & $+0.03\pm0.01$ & $-0.29\pm0.04$ & 0.12 \\
                          & C       & $+0.02\pm0.02$ & $+0.11\pm0.04$ & 0.13 \\
                          & N       & $+0.07\pm0.02$ & $+0.11\pm0.05$ & 0.15 \\
\hline
\end{tabular}
\end{table}

\begin{table*}
\caption{The X-planes for the four elements Mg, Ca, C and N, derived from the combined sample of red sequence galaxies. 
The coefficients are derived from unweighted fits to the model 
${\rm [X/Fe]} =  a_0 + a_1 \log\left(\sigma / 100\,{\rm km\,s}^{-1}\right) + a_2\, {\rm [Fe/H]}$. 
Bold face indicates coefficients with significance level $>$99\%. For the orientation
of the coefficient error ellipses, see Figure~\ref{fig:modellip}. 
For each ratio, we indicate the total dispersion around the fit, the contribution from the errors, and the estimated intrinsic contribution
(all in dex). The scatter calculations include the effects of correlated Fe/H and X/Fe errors, and uncorrelated errors in $\sigma$. 
} 
\label{tab:xplanefit}
\begin{tabular}{lcccccc}
\hline
ratio    &         zero-point & $\sigma$ coeff & Fe/H coeff & total scatter & error & intrinsic \\
& $a_0$ &  $a_1$ &  $a_2$ \\ 
\hline
Mg/Fe    &  {      +0.12$\pm$0.01} & {\bf   +0.35$\pm$0.03} & {\bf   --0.33$\pm$0.04} & 0.07 & 0.06 & 0.04 \\
Ca/Fe    &  {      +0.04$\pm$0.01} & {\bf    +0.15$\pm$0.05} & {\bf   --0.40$\pm$0.05} & 0.10 & 0.08 & 0.05 \\
C/Fe     &  {      +0.06$\pm$0.01} & {\bf    +0.43$\pm$0.04} & {      --0.10$\pm$0.04} & 0.09 & 0.08 & 0.06 \\
N/Fe     &  {      +0.10$\pm$0.02} & {\bf    +0.41$\pm$0.05} & {      --0.07$\pm$0.05} & 0.11 & 0.10 & 0.04 \\
\hline
\end{tabular}
\end{table*}

\begin{table*}
\caption{Extended multi-parametric fits to the abundance ratios, showing the generally non-significant coefficients of age and 
the offset term. The fits are to a model 
${\rm [X/Fe]} =  b_0 + b_1 \log\left(\sigma / 100\,{\rm km\,s}^{-1}\right) + b_2\, {\rm [Fe/H]} +  b_3 \log\left(t_{\rm SSP} / 10\,{\rm Gyr}\right)  + b_4 S$. Here $S=0$ for Coma galaxies and $S=1$ for Shapley galaxies; its coefficient $b_4$ is included to test for
any constant offset between the two samples. 
Bold face indicates coefficients with significance level $>$99\%. The total rms scatter, given in dex, is similar to that of the X-planes in all cases.}\label{tab:unifit}
\begin{tabular}{lcccccc}
\hline
ratio    &         zero-point & $\sigma$ coeff & Fe/H coeff & Age coeff & offset & rms \\
& $b_0$ &  $b_1$ &  $b_2$ & $b_3$ & $b_4$ \\ 
\hline
Mg/Fe    &  {      +0.07$\pm$0.02} & {\bf   +0.31$\pm$0.04} & {\bf   --0.37$\pm$0.05} & {      --0.02$\pm$0.05} & {      +0.05$\pm$0.02} & 0.07 \\
Ca/Fe    &  {     --0.05$\pm$0.03} & {\bf   +0.22$\pm$0.06} & {\bf   --0.54$\pm$0.06} & {\bf      --0.23$\pm$0.07} & {      +0.06$\pm$0.03} & 0.10 \\ 
C/Fe     &  {      +0.07$\pm$0.03} & {\bf   +0.43$\pm$0.06} & {      --0.09$\pm$0.06} & {      +0.01$\pm$0.07} & {      --0.01$\pm$0.03} & 0.09 \\
N/Fe     &  {      +0.11$\pm$0.04} & {\bf   +0.40$\pm$0.07} & {      --0.06$\pm$0.07} & {      +0.01$\pm$0.08} & {      +0.00$\pm$0.03} & 0.11 \\
\hline
\end{tabular}
\end{table*}

\section{Conclusions}\label{sec:concs}

We have analysed the correlations followed by the abundance ratios Mg/Fe, Ca/Fe, C/Fe and N/Fe,
in a sample of red-sequence galaxies over a wide range in mass and metallicity, using data of high signal-to-noise and uniform treatment. 
Using this sample we have discovered new systematic trends in the abundances. In particular, we have demonstrated that in general the 
abundance ratios X/Fe are driven by at least two partially decoupled factors, as evidenced by their planar relationships with Fe abundance and 
velocity dispersion. 

We have shown that the $\alpha$-abundance ratios Mg/Fe and Ca/Fe simultaneously decrease with Fe/H and increase with $\sigma$. 
We interpreted Fe/H dependence as due to a variation in the duration star-formation in different galaxies at given $\sigma$, 
and hence different degrees of Fe enrichment from SNIa\footnote{During final revision of our paper, Graves, Faber \& Schiavon (2008) have confirmed the 
Mg/Fe-vs-Fe/H anti-correlation at fixed $\sigma$ in a sample derived from the Sloan Digital Sky Survey.}.
For both C/Fe and N/Fe, no correlation with Fe/H is observed at fixed $\sigma$. This absence of 
correlation is consistent with these elements being produced primarily by low/intermediate mass stars, 
and hence on a similar time-scale to 
the Fe enrichment.  
We find a striking quantitative similarity in the X/Fe vs Fe/H trends (at fixed $\sigma$), 
with the equivalent trends for galactic stars, 
which suggests a high degree of regularity in the chemical enrichment history of galaxies. 

After allowing for the Fe/H effect, the Mg/Fe, C/Fe and N/Fe ratios all increase with $\sigma$, with a similar slope. 
Coupled with the very different production sites and time-scales for Mg versus C and N, this 
is a powerful constraint on the mechanism responsible
for the $\sigma$ correlations. Several traditional explanations, such as a systematic variation of star-formation time-scale with
$\sigma$ (e.g. Thomas et al. 2005), seem to be disfavoured at face value, because C and N are produced by low/intermediate-mass stars, 
and so should have time-scales similar to Fe. 
The shallow $\sigma$ dependence of Ca/Fe, in comparison to Mg/Fe, remains anomalous, as does the significant additional dependence on
galaxy age, which is found only for Ca. 

In this paper, we have introduced a new diagnostic tool for describing abundance ratios in galaxy samples, and made an initial exploration
of the results obtained for four light elements. Our interpretation of the trends has been necessarily simplistic here, but the discussion 
indicates the greater power of the X-planes, in comparison to the traditional one-parameter scalings typically analysed. We hope that the 
results will stimulate comparisons with more sophisticated chemical evolution models, 
to understand better how the X-planes reflect the star-formation and enrichment history of red-sequence galaxies. 

\section*{Acknowledgments}

RJS was supported for this research under the 
PPARC rolling grant PP/C501568/1 `Extragalactic Astronomy and Cosmology at Durham 2005--2010'.
We are grateful to our collaborators in the Hectospec Coma Survey, Ron Marzke, Ann Hornschemeier and Neal Miller, 
for their contribution to the dataset analysed here, and for their comments on drafts of this paper. 

{}

\appendix

\section{Separate Coma and Shapley fits}\label{sec:app}

In this appendix, we provide results for the single-parameter and X-plane fits obtained for the Shapley sample
and the Coma sample taken separately. 

\begin{figure*}
\includegraphics[angle=270,width=170mm]{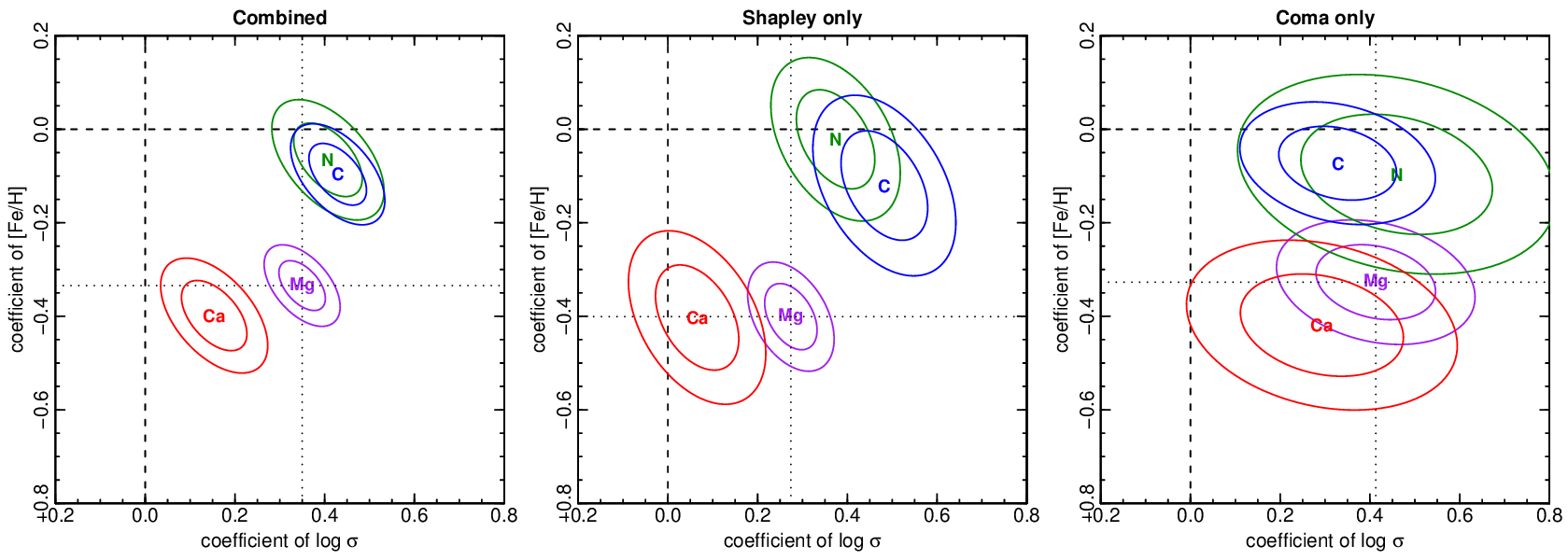}
\caption{Confidence intervals on the X-plane coefficients, for the combined sample and for the two clusters separately.
Ellipses show the 68\% and 95\% confidence intervals on the two slope parameters.}
\label{fig:modellipsplit}
\end{figure*}

Table~\ref{tab:oneparfitsplit} gives the coefficients of the single-parameter fits. The ``Shapey only" and ``Coma only" 
columns correspond to red and blue fit lines in Figures~\ref{fig:mg_fe_corrs}, 
\ref{fig:ca_fe_corrs}, \ref{fig:c_fe_corrs} and \ref{fig:n_fe_corrs}.
In Table~\ref{tab:unifitsplit}, we present the X-pane coefficients for the two cluster samples separately, for
comparison to the combined sample. The confidence intervals on these coefficients are compared in 
Figure~\ref{fig:modellipsplit}. 

These tests demonstrate that the X-plane signal is readily recovered using the Shapley data alone, 
although the coefficients errors are larger, by up to a factor of two. This result further confirms that the 
correlations reported for the combined sample are not driven by unidentified offsets between the two samples. 
Taking only the Coma galaxies, the errors are substantially larger, especially on the $\sigma$ coefficients, 
because the range in $\sigma$ spanned by the Coma dwarfs is very narrow. The range of Fe/H however is fairly large, 
and the Coma sample alone is enough to recover the distinction between Mg and Ca, which show significant 
anti-correlation with Fe/H, and C and N which do not.

\begin{table*}
\caption{The single-parameter fits for X/Fe, for the combined sample, and for the two clusters fit separately.}\label{tab:oneparfitsplit}
\begin{tabular}{lcccccccccccc}
\hline
&& \multicolumn{2}{c}{Combined sample}  && \multicolumn{2}{c}{Shapley only} && \multicolumn{2}{c}{Coma only} \\
Ordinate & X & zero-point & slope &&  zero-point & slope &&  zero-point & slope \\
\hline
$\log\left(\frac{\sigma}{100\,{\rm kms}^{-1}}\right)$   & Mg      &  +0.17$\pm$0.01 &  +0.18$\pm$0.04 &&  +0.17$\pm$0.01 &  +0.14$\pm$0.05 &&  +0.21$\pm$0.04 &  +0.28$\pm$0.12 \\
                                                        & Ca      &  +0.10$\pm$0.02 & --0.05$\pm$0.05 &&  +0.10$\pm$0.02 & --0.07$\pm$0.07 &&  +0.17$\pm$0.06 &  +0.12$\pm$0.16 \\
                                                        & C       &  +0.07$\pm$0.01 &  +0.38$\pm$0.04 &&  +0.07$\pm$0.01 &  +0.44$\pm$0.06 &&  +0.05$\pm$0.03 &  +0.30$\pm$0.08 \\
                                                        & N       &  +0.11$\pm$0.01 &  +0.37$\pm$0.04 &&  +0.11$\pm$0.01 &  +0.37$\pm$0.05 &&  +0.13$\pm$0.05 &  +0.42$\pm$0.14 \\
\hline						     
$\log\left(\frac{L_r}{10^9 L_\odot}\right)$             & Mg      &  +0.11$\pm$0.01 &  +0.05$\pm$0.02 &&  +0.15$\pm$0.04 &  +0.01$\pm$0.03 &&  +0.10$\pm$0.01 &  +0.04$\pm$0.05 \\ 
                                                        & Ca      &  +0.12$\pm$0.01 & --0.03$\pm$0.02 &&  +0.18$\pm$0.04 & --0.08$\pm$0.04 &&  +0.12$\pm$0.02 & --0.01$\pm$0.06 \\
                                                        & C       & --0.07$\pm$0.01 &  +0.13$\pm$0.02 && --0.21$\pm$0.04 &  +0.27$\pm$0.04 && --0.06$\pm$0.01 &  +0.13$\pm$0.04 \\  
                                                        & N       & --0.01$\pm$0.01 &  +0.12$\pm$0.02 && --0.10$\pm$0.04 &  +0.20$\pm$0.04 && --0.01$\pm$0.02 &  +0.09$\pm$0.06 \\ 
\hline
$\log\left(\frac{t_{\rm SSP}}{10\,{\rm Gyr}}\right)$    & Mg      &  +0.18$\pm$0.01 &  +0.28$\pm$0.04 &&  +0.20$\pm$0.01 &  +0.30$\pm$0.05 &&  +0.16$\pm$0.02 &  +0.24$\pm$0.05 \\
                                                        & Ca      &  +0.14$\pm$0.02 &  +0.13$\pm$0.05 &&  +0.11$\pm$0.02 &  +0.10$\pm$0.08 &&  +0.17$\pm$0.02 &  +0.19$\pm$0.07 \\
                                                        & C       &  +0.03$\pm$0.01 &  +0.21$\pm$0.05 &&  +0.08$\pm$0.02 &  +0.25$\pm$0.10 && --0.03$\pm$0.01 &  +0.12$\pm$0.04 \\
                                                        & N       &  +0.07$\pm$0.02 &  +0.19$\pm$0.05 &&  +0.13$\pm$0.02 &  +0.26$\pm$0.08 &&  +0.01$\pm$0.02 &  +0.08$\pm$0.07 \\
\hline
$\left[{\rm Fe/H}\right]$                               & Mg      &  +0.09$\pm$0.01 & --0.16$\pm$0.04 &&  +0.13$\pm$0.01 & --0.24$\pm$0.06 &&  +0.00$\pm$0.02 & --0.28$\pm$0.05 \\
                                                        & Ca      &  +0.03$\pm$0.01 & --0.29$\pm$0.04 &&  +0.05$\pm$0.02 & --0.34$\pm$0.07 && --0.00$\pm$0.03 & --0.34$\pm$0.06 \\
                                                        & C       &  +0.02$\pm$0.02 &  +0.11$\pm$0.04 &&  +0.06$\pm$0.02 &  +0.12$\pm$0.10 && --0.07$\pm$0.02 & --0.04$\pm$0.05 \\
                                                        & N       &  +0.07$\pm$0.02 &  +0.11$\pm$0.05 &&  +0.12$\pm$0.02 &  +0.17$\pm$0.08 && --0.03$\pm$0.03 & --0.06$\pm$0.07 \\
\hline						     
\end{tabular}
\end{table*}

\begin{table*}
\caption{The X-plane coefficients for the combined sample, and for the Shapley and Coma samples separately. 
Bold face indicates coefficients with significance level $>$99\%.}\label{tab:unifitsplit}
\begin{tabular}{lccccccccc}
\hline
&& \multicolumn{2}{c}{Combined sample} && \multicolumn{2}{c}{Shapley only} && \multicolumn{2}{c}{Coma only} \\
ratio  && $\sigma$ coeff & Fe/H coeff && $\sigma$ coeff & Fe/H coeff && $\sigma$ coeff & Fe/H coeff \\
&& $a_1$ &  $a_2$  && $a_1$ &  $a_2$  && $a_1$ &  $a_2$  \\
\hline
Mg/Fe   && {\bf   +0.35$\pm$0.03} & {\bf   --0.33$\pm$0.04} && {\bf   +0.27$\pm$0.04} & {\bf   --0.40$\pm$0.05} &&  {\bf   +0.41$\pm$0.09} & {\bf   --0.33$\pm$0.05} \\
Ca/Fe   && {      +0.15$\pm$0.05} & {\bf   --0.40$\pm$0.05} && {      +0.07$\pm$0.06} & {\bf   --0.40$\pm$0.07} &&  {      +0.29$\pm$0.12} & {\bf   --0.42$\pm$0.07} \\ 
C/Fe    && {\bf   +0.43$\pm$0.04} & {      --0.10$\pm$0.04} && {\bf   +0.48$\pm$0.06} & {      --0.12$\pm$0.08} &&  {\bf   +0.33$\pm$0.09} & {      --0.07$\pm$0.05} \\
N/Fe    && {\bf   +0.41$\pm$0.05} & {      --0.07$\pm$0.05} && {\bf   +0.37$\pm$0.06} & {      --0.02$\pm$0.07} &&  {      +0.46$\pm$0.14} & {      --0.10$\pm$0.08} \\ 
\hline
\end{tabular}
\end{table*}

\label{lastpage}
\end{document}